\documentclass[11pt,a4paper]{article}
\usepackage{jheppub}

\usepackage[latin1]{inputenc} 
\usepackage[T1]{fontenc}
\usepackage[safe]{textcomp}
\usepackage{lmodern} 
\usepackage{epsfig}
\usepackage{float}
\usepackage{amstext}  
\usepackage{amsfonts}
\usepackage{amssymb} 
\usepackage{amsmath}
\usepackage{amsthm}
\usepackage{bm}       
\usepackage[bottom]{footmisc} 
\usepackage{array}                
\usepackage{xcolor} 
\usepackage{framed}
\usepackage{slashed}
\usepackage[absolute]{textpos}
\usepackage{axodraw4j}
\usepackage{dsfont}
\usepackage{multirow}

\usepackage{appendix}
\usepackage{listings}
\usepackage{enumerate}     
\usepackage[bottom]{footmisc} 
\usepackage{array}           
\usepackage{parskip}        
\usepackage{xcolor} 
\usepackage{color}
\usepackage{framed}
\usepackage{fancyhdr}
\usepackage{subfig}

\usepackage{tikz}
\usepackage[absolute]{textpos}
\usepackage{pstricks}
\usepackage{axodraw4j}
\usepackage{graphicx}

\newcommand{\f}[2]{\frac{#1}{#2}}
\newcommand{\sss}[1]{\scriptscriptstyle{#1}}
\newcommand{\ssst}[1]{\scriptscriptstyle{\text{#1}}}
\newcommand{\vv}[2]{\left( \begin{array}{c} #1 \\ #2  \end{array} \right)}
\newcommand{\bea}{\begin{eqnarray}}
\newcommand{\eea}{\end{eqnarray}}
\newcommand{\be}{\begin{equation}}
\newcommand{\ee}{\end{equation}}
\newcommand{\ba}{\begin{align}}
\newcommand{\ea}{\end{align}}
\newcommand{\beas}{\begin{eqnarray*}}
\newcommand{\eeas}{\end{eqnarray*}}
\newcommand{\bes}{\begin{equation*}}
\newcommand{\ees}{\end{equation*}}
\newcommand{\bas}{\begin{align*}}
\newcommand{\eas}{\end{align*}}
 \newcommand{\rig}{\rightarrow}

\newcommand{\eps}{{\varepsilon}}
\newcommand{\cd}{{\cdot}} 
\newcommand{\cf}{C_{\scriptscriptstyle{F}}} 
\newcommand{\ca}{C_{\scriptscriptstyle{A}}}
\newcommand{\tr}{T_{\scriptscriptstyle{F}}}
\newcommand{\dF}{N_{\scriptscriptstyle{c}}}
\newcommand{\dR}{d_{\scriptscriptstyle{R}}}
\newcommand{\Ng}{n_{\scriptscriptstyle{g}}}

\newcommand{\Nf}{n_{\scriptscriptstyle{f}}}
\newcommand{\gs}{g_{\scriptscriptstyle{s}}}
\newcommand{\yt}{y_{\scriptscriptstyle{t}}}

\newcommand{\gb}{g_1}
\newcommand{\gw}{g_2}
\newcommand{\gaf}{\gamma_{\scriptscriptstyle{5}}}
\newcommand{\als}{\alpha_{\scriptscriptstyle{s}}}

\newcommand{\lb}{\left(}
\newcommand{\rb}{\right)}

\newcommand{\msbar}{$\overline{\text{MS}}$}

\newcommand{\dFFf}{d_{\scriptscriptstyle{F}}^{abcd}d_{\scriptscriptstyle{F}}^{abcd}}
\newcommand{\dFAf}{d_{\scriptscriptstyle{F}}^{abcd}d_{\scriptscriptstyle{A}}^{abcd}}

\definecolor{bluemar}{rgb}{0,0,.5}
\definecolor{redmar}{rgb}{.8,0,0}
\definecolor{greenmar}{rgb}{0,.5,0}

\newcommand{\LL}{{\cal L}}
\newcommand{\D}{{\cal D}}

\newcommand{\bigrint}{%
\hbox to 2.92em{\hss\scalebox{1.1}[1] {\rotatebox[origin=c]{15}{$\displaystyle\int$}}\hss}}

\newcommand{\rint}{\!\!\!\!\!\!\!\mathop{\bigrint}\displaylimits\!\!\!\!\!\!\!}
\newcommand{\prd}{\partial}

\catcode`\@=11
\def\slash{\mathpalette\make@slash}
\def\make@slash#1#2{\setbox\z@\hbox{$#1#2$}%
  \hbox to 0pt{\hss$#1/$\hss\kern-\wd0}\box0}
\catcode`\@=12 

\newcommand{\FMslash}{\slash}

\newcommand{\beq}{\begin{equation}}
\newcommand{\eeq}{\end{equation}}

\allowdisplaybreaks
\newcommand{\ice}[1]{\relax}
\title{Leading QCD-induced four-loop contributions to the $\beta$-function of the Higgs self-coupling in the SM and vacuum stability}
\author[a]{K.~G.~Chetyrkin, }
\author[b]{M.~F.~Zoller}
\affiliation[a]{Institut f\"ur Theoretische Teilchenphysik, Karlsruhe
  Institute of Technology (KIT), Germany}
  \affiliation[b]{Institut f\"ur Physik, University of Zurich (UZH), Switzerland}

\emailAdd{konstantin.chetyrkin@kit.edu}
\emailAdd{zoller@physik.uzh.ch}

\abstract{We present analytical results for the leading top-Yukawa and QCD contribution to the $\beta$-function 
for the Higgs self-coupling $\lambda$ of the Standard Model at four-loop level, namely the part $\propto \yt^4 \gs^6$ independently confirming a result given in \cite{Martin:2015eia}.
We also give the contribution $\propto \yt^2 \gs^6$
of the anomalous dimension of the Higgs field as well as the terms $\propto \yt \gs^8$ to the top-Yukawa $\beta$-function which can also be derived from the anomalous dimension
of the top quark mass. We compare the results with the RG functions of the correlators of two and four scalar currents in pure QCD and find a new relation
between the anomalous dimension $\gamma_0$ of the QCD vacuum energy and the anomalous dimension $\gamma_m^{SS}$ appearing in the RG equation of the correlator of two scalar currents.
Together with the recently computed top-Yukawa and QCD contributions
to $\beta_{\gs}$ \cite{Bednyakov:2015ooa,Zoller:2015tha} the $\beta$-functions presented here constitute the leading four-loop contributions to the evolution of the Higgs self-coupling.
A numerical estimate of these terms at the scale of the top-quark mass is presented as well as an analysis of the impact
on the evolution of $\lambda$ up to the Planck scale and the vacuum stability problem.
}

\arxivnumber{}
\subheader{TTP16-008, ZU-TH-3/16}

\begin{document}
\maketitle

\section{Introduction}

The evolution of the Higgs self-coupling and of the Higgs field are important ingredients for the Renormalization Group (RG)
improved Higgs potential and the study of vacuum stability. 
A precise determination of the Higgs self-coupling in the Standard Model extended up to the Planck scale is important
because this parameter is close to zero at the Planck scale and the question whether the SM vacuum state is stable or not
can only be answered definitively by reducing the uncertainties.\footnote{During the last years many detailed studies of the vacuum stability issue in the SM have been performed
\cite{Bezrukov:2009db,Holthausen:2011aa,EliasMiro:2011aa,Xing:2011aa,Bezrukov:2012sa,Degrassi:2012ry,Chetyrkin:2012rz,Zoller:2012cv,
Masina:2012tz,Zoller:2014cka,Zoller:2014xoa,Zoller:2013mra,Buttazzo:2013uya,Bednyakov:2015sca} following the original ideas of 
\cite{Krasnikov:1978pu,Politzer:1978ic,Hung:1979dn}. A recent extension to the MSSM can be found in \cite{Bobrowski:2014dla}.}
The largest source of uncertainty is the experimentally
measured top mass $M_{\ssst{t}}$. At a future linear $e^+e^-$ collider it could, however, be measured with a precision
which matches that of the theory input to the vacuum stability analysis (see Fig.~5 in \cite{Zoller:2014cka}).

On the theory side there are three sources of uncertainty. The first is the difference between the effective Higgs
potential $V_{\ssst{eff}}(\Phi_{\ssst{cl}})$ \cite{PhysRevD.7.1888} and the approximation of the RG-improved potential 
\be
V_{\ssst{RG}}(\Phi_{\ssst{cl}}):=
\lambda(t)\left[ \Phi_{\ssst{cl}}\,\cd\, \text{exp}\lb-\f{1}{2}\int\limits_0^t\!\mathrm{d}t'\gamma_{\sss{\Phi}}(t')\rb \right]^4,\;
t:=\ln\lb\f{\Phi_{\ssst{cl}}^2}{\mu_0^2}\rb,
\label{VRGIdef}
\ee
where \mbox{$\Phi_{\ssst{cl}}=\langle 0 |\Phi|0 \rangle$} is the classical field strength of the scalar SU(2) doublet $\Phi$, $\gamma_{\sss{\Phi}}$
the anomalous dimension of $\Phi$ and $\mu_0$ the scale where we start the evolution of fields and couplings, e.~g.~$\mu_0=M_{\ssst{t}}$. 
This uncertainty is negligible
at large values of $\Phi_{\ssst{cl}}$, e.~g.~close to the Planck scale \cite{Cabibbo:1979ay,Sher:1988mj,Lindner:1988ww,Ford:1992mv,Altarelli1994141}.
In this approximation the SM vacuum is stable up to the scale $\Lambda \sim M_{\ssst{Planck}}$ if $\lambda(\mu)>0$ for $\mu \leq \Lambda$. 

Another source of theoretical uncertainties is the matching of experimental parameters, e.~g.~$M_{\ssst{t}}, \als(M_{\ssst{Z}}), M_{\ssst{H}}$,
to the parameters of the SM Lagrangian, $\yt(\mu_0), \gs(\mu_0), \lambda(\mu_0),\ldots$ at some initial scale $\mu_0$
renormalized in the \msbar-scheme. State of the art is the full numerical two-loop matching \cite{Buttazzo:2013uya,Kniehl:2015nwa}.
In order to improve precision here three-loop calculations might be attempted and different mass definitions than
the pole mass could be used for the top mass.

In this paper we improve on the third source of uncertainty, namely we increase the precision in the $\beta$-functions, calculated in the \msbar-scheme.
The $\beta$-function for a coupling $X$ is defined as
\be
\beta_{\sss{X}}(X,X_1,X_2,\ldots)=\mu^2\f{d X}{d \mu^2}    =\sum \limits_{n=1}^{\infty} \f{1}{(16\pi^2)^{n}}\,\beta_{\sss{X}}^{(n)}
\ee
and the anomalous dimension of a field $f$ as
\be
\gamma^{\sss{f}}_2(X,X_1,X_2,\ldots)=-\mu^2\f{d \text{ln} Z_{\sss{f}}^{-1}}{d \mu^2}    =
\sum \limits_{n=1}^{\infty} \f{1}{(16\pi^2)^{n}}\,\gamma_2^{{\sss{f}}\,(n)}
{},
\ee
where $Z_{\sss{f}}$ is the field strength renormalization constant, where \mbox{$X,X_1,X_2,\ldots$} are the 
couplings of the theory which we want to include in the analysis.

The RG functions of the SM were computed at three-loop accuracy during the last years \cite{PhysRevLett.108.151602,Mihaila:2012pz,Bednyakov:2012rb,Chetyrkin:2012rz,
Chetyrkin:2013wya,Bednyakov:2012en,Bednyakov:2013eba,Bednyakov:2013cpa}. The four-loop $\beta$-function for the strong coupling $\gs$ was first computed in pure 
QCD \cite{4loopbetaqcd,Czakon:2004bu} and recently extended to the gaugeless limit of the SM, namely to include the dependence on the top-Yukawa coupling $\yt$ and 
the Higgs self-coupling $\lambda$ \cite{Bednyakov:2015ooa,Zoller:2015tha}. The leading four-loop contribution to the Higgs self-coupling $\beta$-function was first
presented in \cite{Martin:2015eia} and is independently confirmed in this paper.
%

The paper is structured as follows: In the next section we briefly describe the technical details of the calculation. 
Then the leading four-loop terms for $\beta_{\lambda}$, $\beta_{\yt}$ and $\gamma_2^{\sss{\Phi}}$ are given 
and the relevance of the four-loop terms numerically determined at the scale of the top quark mass. 
Finally, we investigate the impact of the new contributions on the evolution of $\lambda$ in order to estimate
the uncertainty reduction due to four-loop $\beta$-functions.

\section{Technicalities}

\subsection{The Model: QCD plus minimal top-Yukawa contributions}
For this calculation we start with the SM Lagrangian in the broken phase where
\be
\Phi=\vv{\Phi_1}{\Phi_2} \rig \vv{\Phi^+}{\f{1}{\sqrt{2}}\lb v+H-i\chi \rb}. \label{SSBPhi}
\ee
The UV renormalization constants in the \msbar-scheme do not depend on masses and are the same as in the unbroken phase.
Hence we can use all renormalization constants determined up to three-loop level in previous calculations \cite{Chetyrkin:2012rz} and
set all masses to zero.
There is no $\gaf$ in the $t\bar{t}H$-vertex $\propto \yt$ as opposed to the $t\bar{t}\Phi_2$-vertex of the unbroken phase.
$\gaf$ now only appears in the Yukawa vertices with $\chi$ and $\Phi^\pm$. As at low scales (where we start the evolution of couplings and fields) 
the strong coupling is the largest we take as the leading contribution to the vertex and self-energy corrections those where $H$ only appears
as an external field. This means that no Higgs or Goldstone propagators appear. The electroweak gauge-couplings as well as $\lambda$ and all
Yukawa couplings except $\yt$ are neglected.
For the top-Yukawa vertex and the top self-energy these are the pure QCD corrections (see Fig.~\ref{diasZttH}).
For the Higgs self-energy and the quartic Higgs-vertex these are gluon insertions into the one-loop diagram $\propto \yt^2$ (see Fig.~\ref{diasZHH} (a))
and $\propto \yt^4$ (see Fig.~\ref{diasZHHHH} (a))
as well as diagrams with two fermion loops (see Fig.~\ref{diasZHH} (b) and Fig.~\ref{diasZHHHH} (b)). 
Thus we get the four-loop contributions which are numerically most significant to the evolution of $\lambda$
avoiding $\gaf$ and its treatment in $D\neq 4$ dimensions completely.

\begin{figure}[h!]\begin{center}
  \begin{tabular}{cc}
 \scalebox{0.8}{   \begin{picture}(140,100) (0,0)
    \SetWidth{0.5}
    \SetColor{Black}
    \DashLine(70,95)(70,75){4}
    \ArrowLine(70,75)(135,0)
    \ArrowLine(5,0)(70,75)
    \Gluon(93,50)(47,50){4}{6}
    \Gluon(16,12)(124,12){4}{14}
    \Gluon(60,15)(60,46){4}{4}
    \Gluon(80,15)(80,46){4}{4}
    \Vertex(60,15){2}
    \Vertex(80,15){2}
    \Vertex(60,46){2}
    \Vertex(80,46){2}
    \Text(10,-15)[lb]{\Large{\Black{$t$}}}
    \Text(125,-15)[lb]{\Large{\Black{$t$}}}
    \Text(75,90)[lb]{\Large{\Black{$H$}}}
    \Text(5,75)[lb]{\Large{\Black{(a)}}}
  \end{picture} }\qquad\qquad\qquad
 &   \scalebox{0.8}{  \begin{picture}(140,100) (0,0)
    \SetWidth{0.5}
    \SetColor{Black}
    \ArrowArc(70,30)(40,0,180)
     \ArrowLine(110,30)(30,30)
    \Gluon(30,30)(30,0){4}{4}
    \Gluon(55,30)(55,0){4}{4}
    \Gluon(85,30)(85,0){4}{4}
    \Gluon(110,30)(110,0){4}{4}
    \ArrowLine(5,0)(30,0)
    \ArrowLine(30,0)(110,0)
    \ArrowLine(110,0)(135,0)
    \Text(10,-15)[lb]{\Large{\Black{$t$}}}
    \Text(125,-15)[lb]{\Large{\Black{$t$}}}
    \Text(5,75)[lb]{\Large{\Black{(b)}}}
  \end{picture}}\\[4ex]
\end{tabular}\end{center}
\caption{Some diagrams contributing to the Yukawa correction (a) and the top quark self-energy (b)}
\label{diasZttH}
\end{figure}
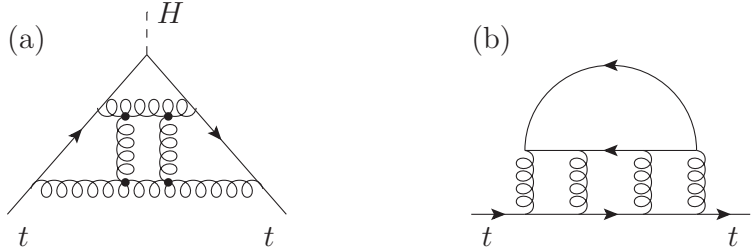

\begin{figure}[h!] \begin{center}
  \begin{tabular}{cc}
   \scalebox{0.8}{ \begin{picture}(140,130) (0,0)
    \SetWidth{0.5}
    \SetColor{Black}
    \DashLine(5,0)(15,10){4}
    \DashLine(115,10)(125,0){4}
    \DashLine(5,110)(15,100){4}
    \DashLine(115,100)(125,110){4}
    \ArrowLine(15,10)(115,10)
    \ArrowLine(115,100)(15,100)
    \ArrowLine(15,100)(15,10)
    \ArrowLine(115,10)(115,100)
    \Gluon(45,100)(45,10){4}{12}
    \Gluon(85,10)(85,100){4}{12}
    \Gluon(48,55)(82,55){4}{4}
    \Vertex(48,55){2}
    \Vertex(82,55){2}
    \Text(10,-15)[cb]{\Large{\Black{$H$}}}
    \Text(120,-15)[cb]{\Large{\Black{$H$}}}
    \Text(10,125)[ct]{\Large{\Black{$H$}}}
    \Text(120,125)[ct]{\Large{\Black{$H$}}}
    \Text(70,-30)[cb]{\Large{\Black{(a)}}}
  \end{picture} }\qquad\qquad\qquad
 &  \scalebox{0.8}{    \begin{picture}(140,130) (0,0)
    \SetWidth{0.5}
    \SetColor{Black}
    \DashLine(5,0)(15,10){4}
    \DashLine(115,10)(125,0){4}
    \DashLine(5,110)(15,100){4}
    \DashLine(115,100)(125,110){4}
    \ArrowLine(15,10)(45,10)
    \ArrowLine(45,100)(15,100)
    \Gluon(45,10)(85,10){4}{5}
    \Gluon(45,100)(85,100){4}{5}
    \ArrowLine(85,10)(115,10)
    \ArrowLine(115,100)(85,100)
    \ArrowLine(15,100)(15,10)
    \ArrowLine(115,10)(115,100)
    \ArrowLine(45,55)(45,10)
    \ArrowLine(85,10)(85,55)
    \ArrowLine(45,100)(45,55)
    \ArrowLine(85,55)(85,100)
    \Gluon(45,55)(85,55){4}{5}
    \Text(10,-15)[cb]{\Large{\Black{$H$}}}
    \Text(120,-15)[cb]{\Large{\Black{$H$}}}
    \Text(10,125)[ct]{\Large{\Black{$H$}}}
    \Text(120,125)[ct]{\Large{\Black{$H$}}}
    \Text(70,-30)[cb]{\Large{\Black{(b)}}}
  \end{picture} }\\[6ex]
\end{tabular} \end{center}
\caption{Some diagrams contributing to the quartic Higgs self-interaction}
\label{diasZHHHH}
\end{figure}
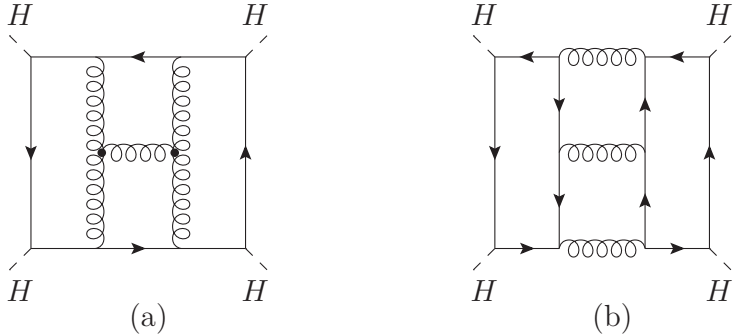

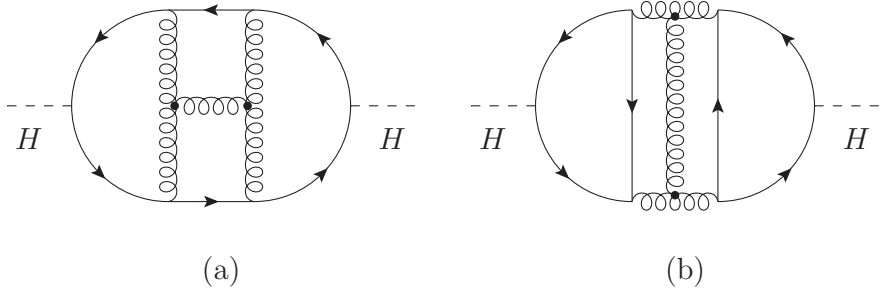
\begin{figure}[h!] \begin{center}
  \begin{tabular}{cc} \scalebox{0.8}{
    \begin{picture}(140,130) (0,0)
    \SetWidth{0.5}
    \SetColor{Black}
    \DashLine(-30,55)(0,55){4}
    \DashLine(130,55)(160,55){4}
    \ArrowLine(45,10)(85,10)
    \ArrowLine(85,100)(45,100)
    \ArrowArc(45,55)(45,90,180)
    \ArrowArc(45,55)(45,180,270)
    \ArrowArc(85,55)(45,270,360)
    \ArrowArc(85,55)(45,0,90)
    \Gluon(45,100)(45,10){4}{12}
    \Gluon(85,10)(85,100){4}{12}
    \Gluon(48,55)(82,55){4}{4}
    \Vertex(48,55){2}
    \Vertex(82,55){2}
    \Text(-20,35)[cb]{\Large{\Black{$H$}}}
    \Text(150,35)[cb]{\Large{\Black{$H$}}}
    \Text(70,-30)[cb]{\Large{\Black{(a)}}}
  \end{picture} }\qquad\qquad\qquad 
 &  \scalebox{0.8}{    \begin{picture}(140,130) (0,0)
    \SetWidth{0.5}
    \SetColor{Black}
    \DashLine(-30,55)(0,55){4}
    \DashLine(130,55)(160,55){4}
    \ArrowArc(45,55)(45,90,180)
    \ArrowArc(45,55)(45,180,270)
    \ArrowArc(85,55)(45,270,360)
    \ArrowArc(85,55)(45,0,90)
    \ArrowLine(45,100)(45,10)
    \ArrowLine(85,10)(85,100)
    \Gluon(45,10)(85,10){4}{5}
    \Gluon(85,100)(45,100){4}{5}
    \Gluon(65,13)(65,97){4}{12}
    \Vertex(65,97){2}
    \Vertex(65,13){2}
    \Text(-20,35)[cb]{\Large{\Black{$H$}}}
    \Text(150,35)[cb]{\Large{\Black{$H$}}}
    \Text(70,-30)[cb]{\Large{\Black{(b)}}}
  \end{picture}}\\[6ex]
\end{tabular} \end{center}
\caption{Some diagrams contributing to the Higgs self-energy}
\label{diasZHH}
\end{figure}

We compute the field strength renormalization constants $Z_2^{\sss{(HH)}}$ from the Higgs and $Z_2^{\sss{(\bar{t}t)}}$ from the top self-energies.
The quartic Higgs-vertex is renormalized with $Z_1^{\sss{(4H)}}$ and the top-Yukawa-vertex with $Z^{\sss{(\bar{t}tH)}}_1$.\footnote{\parbox{\textwidth}{In the notation of \cite{Chetyrkin:2012rz} these renormalization constants are
\mbox{$Z_2^{\sss{(HH)}}=Z_2^{\sss{(2\Phi)}}$,} \mbox{$Z_2^{\sss{(\bar{t}t)}}=Z_2^{\sss{(2t)}}$,} \mbox{$Z_1^{\sss{(4H)}}=Z_1^{\sss{(4\Phi)}}$} and 
\mbox{$Z^{\sss{(\bar{t}tH)}}_1=Z^{\sss{(tb\Phi)}}_1$.}}}

>From these we compute
\be \delta\! Z_\lambda= \f{\lambda-\delta\! Z_1^{\sss{(4H)}}}{\left(Z_2^{\sss{(HH)}}\right)^{2}}-\lambda \ee
and
\be 
Z_{\yt}=\f{Z^{\sss{(\bar{t}tH)}}_1}{Z^{\sss{(\bar{t}t)}}_{2,}\sqrt{Z_2^{\sss{(HH)}}}}
{}. 
\ee
All divergent integrals are regularized in $D=4-2 \eps$ space time dimensions and the renormalization constants are defined as $Z=1+\delta Z$
in the \msbar-scheme.

\subsection{Calculation with massive tadpole integrals}
For the computation of the four-loop terms we use the setup described in detail in  \cite{Zoller:2015tha}.
The generation of all necessary Feynman diagrams was done with QGRAF \cite{QGRAF}. 
The C++ programs Q2E and EXP \cite{Seidensticker:1999bb,Harlander:1997zb} are then used to identify the topology of the diagram. 
The Taylor expansion in external momenta, the fermion traces and the insertion of counterterms in lower loop diagrams was performed with
FORM \cite{Vermaseren:2000nd,Tentyukov:2007mu}. All colour factors were computed with the \mbox{FORM} package 
COLOR \cite{COLOR}.

In the momentum space part of the diagrams we introduce the same auxiliary mass parameter $M^2$ 
in every propagator denominator.
The self-energy diagrams are then expanded to second order in the external momentum $q$ after applying a projector $\propto \slashed{q}$
to the top self-energy diagrams and taking the trace over the external fermion line. 
Then we divide by $q^2$ before $q$ is set to zero. In all vertex correction diagrams we can set $q \to 0$ from
the beginning. This is allowed as \msbar{} renormalization constants do not depend on external momenta. 
After this we are left with tadpole integrals. 
Subdivergences $\propto M^2$ are canceled by counterterms
\be \begin{split}
\f{M^2}{2}\delta\!Z_{\sss{M^2}}^{(2g)}\,A_\mu^a A^{a\,\mu}
\end{split} \ee
computed from and inserted in lower loop diagrams.
This is the same method for computing UV renormailzation constants as in our previous calculations 
\cite{Chetyrkin:2012rz,Chetyrkin:2013wya,Zoller:2015tha}.
It was first introduced in \cite{Misiak:1994zw} and then further developed in \cite{beta_den_comp}.
A detailed explanation of the calculation of Z-factors with an auxiliary mass can be found in \cite{Zoller:2014xoa}.

Up to three-loop order the tadpole integrals were computed with the \mbox{FORM}-based package \mbox{MATAD}\cite{MATAD}.
The four-loop tadpoles are reduced to Master integrals using FIRE \cite{Smirnov:2008iw,Smirnov:2014hma}. 
The needed four-loop Master integrals can be found in \cite{Czakon:2004bu}.

\section{Analytical Results \label{res:beta}}
In this section we give our results which can be found in machine readable format on\\
\texttt{\bf http://www-ttp.particle.uni-karlsruhe.de/Progdata/ttp16/ttp16-008/}\\
For a gerneric SU($\dF$) gauge group the colour factors are expressed
through the quadratic Casimir operators $\cf$ and $\ca$ of the 
fundamental and the adjoint representation of the corresponding Lie algebra.
The dimension of the fundamental representation is called $\dR$. The adjoint representation has dimension $\Ng$ and the trace
$\tr$ is defined by \mbox{$\tr \delta^{ab}=\textbf{Tr}\lb T^a T^b\rb$}  
with the group generators $T^a$ of the fundamental representation. 
Higher order invariants are constructed from the symmetric tensors
\bea
d_{\scriptscriptstyle{F}}^{abcd} &=& \f{1}{6} \text{Tr} 
\lb T^a T^b T^c T^d + T^a T^b T^d T^c + T^a T^c T^b T^d \right. \nonumber \\
&+& \left. T^a T^c T^d T^b + T^a T^d T^b T^c + T^a T^d T^c T^b \rb{}.
\eea
from the generators of the fundamental representation and analogously $d_{\scriptscriptstyle{A}}^{abcd}$ from the generators of the adjoint representation.
The combinations needed and their SU($\dF$) values are
\be
\dFFf = \Ng\lb\frac{\dF^4-6\dF^2+18}{96\dF^2}\rb,\;\;\;\;
\dFAf = \Ng\lb\frac{\dF(\dF^2+6)}{48}\rb.
\ee
Furthermore for SU($\dF$) we have \be \dR=\dF,\;\;\;\; \tr = \f{1}{2},\;\;\;\; \cf = \f{\dF^2-1}{2\dF},\;\;\;\; \ca = \dF,\;\;\;\; \Ng = \dF^2-1{}.\ee
The number of active fermion flavours is denoted by $\Nf$. The leading four-loop contributions to the $\beta$-functions
for the Higgs self-coupling and the top-Yukawa coupling are found to be
\be
\begin{split}
\beta_{\sss{\lambda}}^{(4)}= &
        \yt^4 \gs^6  \,\dR\, \left\{
        \cf^3   \lb - \f{2942}{3}  
          + 160 \zeta_{5} 
           + 288 \zeta_{4} 
           + 48 \zeta_{3} \rb  \right. \\ &\left.
       +\tr \cf^2 \lb  - 64  +\Nf\lb
          + \f{562}{3}  
          - 160 \zeta_{4} 
          + \f{32}{3} \zeta_{3} \rb \rb  \right. \\ &\left.
        +\ca \cf^2 \lb \f{3584}{3} 
          + 720 \zeta_{5} 
          + 32 \zeta_{4} 
          - \f{3304}{3} \zeta_{3} \rb  \right. \\ &\left.
        +\ca \tr \cf  \lb \f{5888}{9} 
          - 160 \zeta_{5}
          + 352 \zeta_{3} +\Nf \lb
          - \f{2644}{243}  
          + 128 \zeta_{4} 
          + 16 \zeta_{3}  \rb\rb  \right. \\ &\left.
         +\ca^2 \cf  \lb - \f{121547}{243}  
          - 520 \zeta_{5} 
          - 88 \zeta_{4}
          + \f{1880}{3} \zeta_{3} \rb  \right. \\ &\left.
         +\tr^2 \cf \lb - \f{256}{9} \Nf  + \Nf^2\lb
          - \f{128}{3} \zeta_{3}  
          + \f{10912}{243} \rb\rb
          \right\}\\ &+ \mathcal{O}(\yt^6)+ \mathcal{O}(\lambda)+ \mathcal{O}(\gw)+ \mathcal{O}(\gb)
\end{split}
\label{4lbetalambda}
\ee
in agreement\footnote{Note that in \cite{Martin:2015eia} the definition $\beta_{\lambda}^{\ssst{S.M.}}=\mu\f{d\lambda}{d\mu}=2\beta_\lambda$
for the $\beta$-function is used.} with eq.~(4.32) of \cite{Martin:2015eia} and
\be
\begin{split}
\beta_{\sss{\yt}}^{(4)}= &
\yt \gs^8   \left\{
           \f{\dFAf}{\dR} \lb 32
           - 240 \zeta_{3} \rb           
           +\Nf \f{\dFFf}{\dR} \lb - 64 
           + 480 \zeta_{3} \rb  \right. \\ &\left.            
          +\cf^4 \lb \f{1261}{8} 
          + 336 \zeta_{3} \rb          
          - \ca \cf^3 \lb \f{15349}{12} +
           316 \zeta_{3} \rb  \right. \\ &\left.           
          +\ca^2 \cf^2\lb \f{34045}{36}           
          - 440 \zeta_{5} 
          + 152 \zeta_{3} \rb          
          +\ca^3 \cf \lb - \f{70055}{72}  
          + 440 \zeta_{5} 
           - \f{1418}{9} \zeta_{3} \rb  \right. \\ &\left.           
          +\Nf \tr \cf^3\lb \f{280}{3} 
          + 480 \zeta_{5} 
          - 552 \zeta_{3} \rb           \right. \\ &\left. 
          +\Nf \ca \tr \cf^2\lb \f{8819}{27} 
          - 80 \zeta_{5} 
          + 264 \zeta_{4}  
          - 368 \zeta_{3} \rb  \right. \\ &\left.           
          +\Nf \ca^2 \tr \cf\lb \f{65459}{162}  
          - 400 \zeta_{5} 
           - 264 \zeta_{4} 
           + \f{2684}{3} \zeta_{3} \rb      \right. \\ &\left.      
          + \Nf^2 \tr^2 \cf^2 \lb - \f{304}{27}
          - 96 \zeta_{4} 
          + 160 \zeta_{3} \rb  \right. \\ &\left.           
          +\Nf^2 \ca \tr^2 \cf \lb - \f{1342}{81} 
          + 96 \zeta_{4} 
          - 160 \zeta_{3} \rb          \right. \\ &\left.  
          +\Nf^3 \tr^3 \cf\lb \f{664}{81} 
          - \f{128}{9} \zeta_{3} \rb     
          \right\}\\ &+ \mathcal{O}(\yt^3)+ \mathcal{O}(\lambda)+ \mathcal{O}(\gw)+\mathcal{O}(\gb).
\end{split}
\label{4lbetayt}
\ee
The leading contribution to the anomalous dimension of the Higgs field (or equivalently the scalar SU(2) 
doublet $\Phi$) 
at four-loop level is given by
\be
\begin{split}
\gamma_2^{\sss{\phi}\,(4)}= &
       \yt^2\gs^6  \,\dR\, \left\{
           \cf^3\lb \f{4651}{12}  
           - 360 \zeta_{5} 
           - 108 \zeta_{4} 
           + 232 \zeta_{3} \rb       \right. \\ &\left.    
         +\ca \cf^2\lb - \f{1282}{3} 
          - 180 \zeta_{5}
          - 78 \zeta_{4} 
          + 518 \zeta_{3}\rb        \right. \\ &\left.  
          + \ca^2 \cf\lb \f{267889}{972}  
          + 180 \zeta_{5} 
          + 66 \zeta_{4} 
          - \f{950}{3} \zeta_{3} \rb    \right. \\ &\left.      
          +\Nf \tr \cf^2\lb - 125 
          + 96 \zeta_{4} 
          + 16 \zeta_{3} \rb          
          - \Nf \ca \tr \cf\lb \f{631}{243} 
          + 72 \zeta_{4}
          + \f{160}{3} \zeta_{3} \rb    \right. \\ &\left.      
          +\Nf^2 \tr^2 \cf \lb - \f{6500}{243} 
         + \f{64}{3} \zeta_{3} \rb         
          \right\}\\ &+ \mathcal{O}(\yt^4)+ \mathcal{O}(\lambda)+ \mathcal{O}(\gw)+ \mathcal{O}(\gb).
\end{split}
\label{4lgammaHH}
\ee
For the lower loop contributions we refer to \cite{Chetyrkin:2012rz,Chetyrkin:2013wya,Bednyakov:2012en,Bednyakov:2013eba,Bednyakov:2013cpa}.

\section{Comparison with available QCD results  }

All diagrams discussed in the previous section are special in one  aspect: they comprise just the minimal number
of non-QCD vertexes, that is  one for $\beta_{y_t}$,  two for  $\gamma_2^{\Phi}$  and
\mbox{four for $\beta_{\lambda}$.}
Even more, these non-QCD vertexes are of one and the same type, namely the
insertion of the {\em scalar} top-quark current $\bar{t}\, t$.
This  means that the corresponding anomalous dimensions should
be related  to some RG functions in pure QCD describing the QCD evolution of the
scalar current(s).  
The  corresponding  ``effective'' Lagrangian  
\beq
{\cal L}_{eff} =  {\cal L}_{QCD} -\frac{y_t}{\sqrt{2}} H \bar{t} t  -
\frac{\lambda}{4} H^4
\eeq
implies, obviously,  the following identities {\bf valid in all orders} in
$\alpha_s$
(dots below  stand for terms which   have a different  dependence on the SM coupling constants than 
$ y_t\, \alpha_s^n$ (first line) and  $ y^2_t\, \alpha_s^n$ (second line)   correspondingly):
\bea
\beta_{\yt} &=& y_t\, \gamma_m(\alpha_s)\  + \   \dots
\label{gm}
{},
\\
\gamma_2^{\sss{\Phi}} &=& \frac{y_t^2}{2}\, \gamma^{SS}_q(\alpha_s)   \ + \ \dots
\label{SS}
{}.
\eea
Here $\gamma_m(\alpha_s)$ is the quark mass anomalous  dimension and the
function   $\gamma^{SS}_q(\alpha_s)$ appears in the evolution equation 
\[
\mu^2 \frac{\mathrm d}{\mathrm{d} \mu^2} \Pi^S = -2\gamma_m \,  \Pi^S +   \gamma^{SS}_q Q^2 + \gamma^{SS}_m m_t^2 
\]
for the scalar correlator ($B$ marking bare quantities)
\beq
\Pi_{\sss{B}}^S(Q^2,m_t)\, = i\,\rint \, \mathrm{d}^Dx\, e^{iqx} \langle 0|T\left[j^{\sss{B}}_s(x)
  j^{\sss{B}}_s(0)\right] |0\rangle, \ \ j^{\sss{B}}_s= \bar{t_{\sss{B}}}
t_{\sss{B}}, \  \  Q^2 = -q^2
{}.
\nonumber
\eeq
which is renormalized as
\beq
\Pi^S(q^2,m_t)\, = Z_m^2\Pi_B^S(q^2,m_t^B) + \lb Z_2^{SS} Q^2 + Z_m^{SS} m_t^2\rb \mu^{-2\eps}
{}.
\eeq
The anomalous dimensions are found to be
\bea
\gamma_m &=& -\f{d\ln Z_m}{d\ln\mu^2}, \\
\gamma^{SS}_q &=& \f{d Z_2^{SS}}{d\ln\mu^2} + \lb 2\gamma_m-\eps\rb Z^{SS}_2,\\
\gamma^{SS}_m &=& \f{d Z_m^{SS}}{d\ln\mu^2} + \lb 4\gamma_m-\eps\rb Z^{SS}_m.
\eea
Needless to say that a comparison with  available results for
$\gamma_m(\alpha_s)$ \cite{Vermaseren:1997fq,Chetyrkin:1997dh} and
$\gamma^{SS}_q(\alpha_s)$ \cite{Chetyrkin:1996sr} confirms the relations
(\ref{gm}) and (\ref{SS}) at four-loop order\footnote{
In fact, both quantities are currently also  known to {\bf five} loops from \cite{Baikov:2014qja,Baikov:2005rw}. 
}.

\newcommand{\re}[1]{(\ref{#1})}
\newcommand{\unl}[1]{\underline{#1}}

Finally, let us consider  in some  detail the last (and somewhat more  complicated)  case, viz.   the  $\beta$-function for  
the Higgs self-coupling $\beta_{\lambda}$.  The corresponding renormalization
constant coincides  up to  a factor $\frac{y_t^4}{4\cdot 4!}$   to that which renormalizes the (1PI) Green's
function of the T-product of {\bf four} scalar currents:
\bea
&{}&(2\pi)^{D}\delta\left( p_1 + p_2 +p_3 + p_4 \right)\, \Gamma(\{p_i\},\alpha_s,m_t,\mu) 
\\
=
&{}&
Z_{m}^4\,Z_2^2 \int\mathrm{d}^4x_1\ldots \mathrm{d}^4x_4\, 
\lb \prod\limits_i e^{i p_i \cdot x_i}\rb \langle 0| T\left[j_s(x_1)\,j_s(x_2)\,j_s(x_3)\,j_s(x_4)\right] |0\rangle
\label{HHHH}
{}.
\eea
With  $m_t \not=0$ we could  nullify all  external momenta $p_i$ in 
eq.~\re{HHHH} and, thus, consider all \mbox{$j_s$-operators} on the rhs of  \re{HHHH} as insertions
of the  scalar quark current    {\em at zero momentum} transfer.

As is well-known such insertions can be  generated by
multiple  differentiations of QCD Green functions wrt a quark mass\footnote{For the Higgs decay via heavy top loops such
  relations have been known as {\em low-energy} theorems for a long time 
\cite{Ellis:1975ap,Shifman:1978zn,Kniehl:1995tn,Chetyrkin:1997un}.}.
This  means that the corresponding anomalous dimensions should
be related  to some  pure QCD RG functions. 
The well-known way to construct the corresponding relations is to use the
(renormalized) Quantum  Action Principle \cite{Lowenstein:1971jk,Lam:1972mb}. 
Let us briefly outline  the main points.  
 
The Quantum Action Principle  relates  properties
of (regularized)  Lagrangian and the full Green's functions. Consider the generating functional
of (connected) Green's functions
\beq
W(J)=\sum\limits_{n=1}^{\infty} \f{1}{n!}\int\!\mathrm{d}^4x_1\ldots\mathrm{d}^4x_n \,
G^{(n)}(x_1,\ldots,x_n)\,J(x_1)\ldots J(x_n) \label{WJexpansion}
\eeq
defined in 
\beq
Z(\LL,J) = e^{iW(J)}=  \rint   \D\,\Phi e^{i S(\Phi) + \int\!\Phi\cdot J \mathrm{d}^4x},
\ \  S(\Phi) = \rint \LL(\Phi) \mathrm{d}^4x
{}.
\label{gen:fun}
\eeq

The Action Principle states (in particular)  that 
\beq
\frac{\prd }{ \prd  \lambda}   W(J) \equiv 
\Biggl(\rint   \D\,\Phi e^{i S(\Phi) +\int\!\Phi\cdot J \mathrm{d}^4x}  \frac{\prd }{ \prd  \lambda} S(\Phi)\Biggr) /  Z(\LL,J)
{},
\eeq
where $\lambda$ is a any parameter in  the Lagrangian $\LL$. The action principle works for
DR Green functions   \cite{Breitenlohner:1977hr}     (modulo axial anomalies).

\begin{figure}[!ht]
\begin{center}
\includegraphics[width=0.56\textwidth]{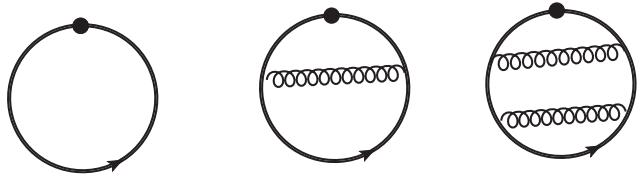} 
\end{center}
\caption{Sample diagrams contributing to the VEV of $Z_m Z_2\,\bar t t(x)$} \label{o2dias}
\end{figure}

An example: the (renormalized) QCD Lagrangian  with $n_l$ massless quarks and
a massive (top) one is customarily written as
\begin{eqnarray}
{\cal L}_{QCD}  = 
&-& \frac{1}{4} Z_3\, ( \partial _{\mu}A_{\nu} -  \partial _{\nu}A_{\mu})^2
- \frac{1}{2}g\, Z_1^{3g} \, ( \partial _{\mu}A^a_{\nu} -  \partial _{\nu}A^a_{\mu})
\,  ( A_{\mu} \times A_{\nu})^a 
\nonumber
\\
&-&
 \frac{1}{4} g^2\, Z^{4g}_1\, ( A_\mu \times A_\nu)^2
\label{lag:2}
+
Z_2 
\sum_{i=1}^{n_l}
\bar \psi_i \,  ( \mathrm{i}  \slash{ \partial } 
 + g Z^{\psi\psi g}_1 Z_2^{-1}\FMslash{A})\, \psi_i
\\
&+&
Z_2 \,
\bar{t}\, ( \mathrm{i}  \slash{ \partial } 
 + g Z^{\psi\psi g}_1 Z_2^{-1}\FMslash{A} - Z_m m_t )\,t
\nonumber
{}.
\end{eqnarray}
With properly chosen renormalization constants $Z_i$
\re{lag:2} should
produce {\em finite} Green's functions.
If one differentiates  $W(J)$   wrt the (renormalized) top quark mass $m_t$ the functional 
\beq
\Biggl(\rint   \D\,\Phi e^{i S(\Phi) + \Phi\cdot J}
 \left(\rint  Z_m Z_2 \, \bar t t(x) d x \right) \, \Biggr)  /  Z(\LL,J)
\label{lag:3}
\eeq should also be finite. However, let us
consdider eq.~\re{lag:3} at $J=0$. It corresponds, obviously, to the VEV of
the operator $Z_m Z_2\,\bar t t(x)$, which is {\em not} finite already
at order $\alpha_s^0$ (a couple of typical diagrams contributing  to \re{lag:3}
at $J=0$ are shown on Fig.~\ref{o2dias}).

This means that our QCD Lagrangian \re{lag:2} is not {\em full}: the term responsible for the renormalization of the 
{\em vacuum energy} is missing. 
The full QCD Lagrangian  reads \cite{Spiridonov:1988md}
\beq
\mathcal{L}_{QCD}^{{\sss{full}}}=\mathcal{L}_{QCD} -  E_0^B, \ \
E_0^B = 
\mu^{-2\epsilon}(E_0(\mu)-Z_0(\alpha_s)m_t^4(\mu)),\qquad \epsilon=(4-D)/2
{},
\eeq
here  $E_0(\mu)$ is the (renormalized) vacuum energy and $Z_0$ is the corresponding
renormalization constant.  

\ice{
Note that the the term ``vacuum energy'' is the right one  as according  to the 
Gell-Mann-Low formula  the expression $\ln G(\LL,J=0)$  which is made finite with the renormalization
constant $E_0$ is 
 equal to 
\[  
\ln \left ( \langle0|\hat{U}(\tau_2,\tau_1) \rangle \right ) = -i\Delta E (\tau_2 - \tau_1) \ \ 
\mbox{in the limit of}  \ \ (\tau_2 - \tau_1) \to \infty
\]
Here $\Delta E = \rho_V  V $  is the shift of the vacuum energy due to interaction, and, thus,
$\rho_V$ is vacuum energy  density.
}

Now, a four-fold differentiation of the generating functional \re{gen:fun}  (with
$\mathcal{L} = \mathcal{L}_{QCD}^{{\sss{full}}}$)
 wrt $m_t$ immediately leads us to the conclusion that
the   combination
\beq
\Gamma(\{p_i\},\alpha_s,m_t,\mu)  - i \, 4! Z_0
\eeq
should be finite. 
As a result, we arrive  at  the following identity {\bf valid in all orders} in $\alpha_s$:
\beq
\beta_{\lambda}   =     y_t^4\,\gamma_0(\alpha_s) \   + \ \dots
\label{be_lam_g0}
{},
\eeq
where the
dots stand for terms which have a different  dependence on the SM coupling constants than 
$ y^4_t\, \alpha_s^n$ and\footnote{We define the QCD  $\beta$-function as 
$ \normalsize \beta(\als) =  \frac{d \als}{d \ln \mu^2}$.
}
\beq
\gamma_0 = (4\gamma_m -\eps)\, Z_0  + (\beta  -\eps)\, \als\frac{\prd Z_0}{\prd \als}
\eeq 
is 
the anomalous dimension of the vaccuum energy
\beq
\frac{d E_0}{d \ln \mu^2} =  \gamma_0(\als)\, m_t^4 
{}.
\eeq

The vacuum anomalous dimension plays an important role in the description of the
renormalization mixing of all three scalar gauge-invariant operators with
(mass)   dimension four:
\bea 
O_1 &=& -\frac{1}{4}(G_{\mu\nu})^2, \ O_2 = m_t \bar t\,t , \  O_3 = m_t^4,
\label{O1_o3}
\\
&{}& 
\mu^2\frac{d}{d\mu^2}O_i = \sum_{j=1}^3 \gamma_{ij} O_j
\label{ops:evol}
{}.
\eea
It was proven in \cite{Spiridonov:1984br,Spiridonov:1988md} that the
matrix  of anomalous  dimensions in \re{ops:evol} reads:
\begin{equation}
\gamma_{ij}=\lb \mu^2\f{d}{d\mu^2} Z_{ik}\rb \lb Z^{-1}\rb_{kj} = 
\left(\begin{array}{ccc}
-\alpha_s\frac{\partial\beta}{\partial\alpha_s}&
-\alpha_s\frac{\partial\gamma_m}{\partial\alpha_s}&
-\alpha_s\frac{\partial\gamma_0}{\partial\alpha_s}\\
0&0&-4\gamma_0\\
0&0&4\gamma_m\\
\end{array}
\right)
\label{mixing}
{}.
\end{equation}

In addition, the two-point scalar correlator \re{SS} at $q=0$ is obviously
related (via the Action Principle) to the four-point correlator \re{HHHH}, as
the latter can be obtained from the former by a double differentiation wrt
$m_t$.  This, in turn, leads to the following remarkable relation (again valid
in all orders in $\alpha_s$):
\beq
\gamma_0 = \frac{1}{12}\gamma^{SS}_m
\label{g0:SS}
{}.
\eeq

Thus, one could compute $\gamma_0$ in a few  different
ways. 
\ice{
Let us, finally, describe the current situation with computing perturbative 
contribitionsto f the (QCD) vacuum anomalous
dimension $\gamma_0$.
Let  discuss separately the  last  case, viz.,  the current status of the (QCD)  vacuum  anomalous dimension
$\gamma_0$.
}


1. Direct renormalization of the vacuum energy diagrams.  This was done for
two and three loops in the  papers \cite{Spiridonov:1988md} and
\cite{Chetyrkin:1994ex} respectively. At four loops it was first found in this way
for a space-time dimension $ D=3 $ \cite{Schroder:2002re,DiRenzo:2004ws}
(in the process of computing the free energy in the effective high temperature
QCD) and (implicitly, via  eq. \re{be_lam_g0}) in the present paper.   

2. By renormalizing the 4-loop scalar correlator \cite{g0as3} in the limit of small
quark mass\footnote{That is  the scalar corelator was expanded at the large
momentum limit  and the $\gamma_m^{SS}$ was found by renormalizing the term of
order $m_t^2/Q^2$.}. The result was later
used in \cite{Chetyrkin:2000zk} to  compute the quartic mass corrections to $R_{had}$ at {${\cal O}(\alpha_s^3)$}.

3. By computing the lowest moment of the scalar correlator  at 4 loops
\cite{Sturm:2008eb}.

4. By  computing the lowest low-energy moment of the axial-vector correlator (related via a
Ward identity to the VEV of the scalar current) \cite{Maier:2009fz}.


Note, finally,  that the last two  evaluations  dealt  with massive tadpole diagrams  and
that  all  calculations of $\gamma_0$ at the four-loop level described above are in
mutual agreement as well as the two results for $\beta_{\lambda}^{(4)}$: the one
displayed in \re{4lbetalambda} and the one obtained via relation \re{be_lam_g0}.

\section{Numerical analysis}

In this section we want to numerically evaluate the results for the $\beta$-functions presented in section \ref{res:beta}. The couplings $\gs, \gw, \gb, \yt$ and  $\lambda$ in the $\overline{\text{MS}}$-scheme
at some fixed scale $\mu_0$ can be computed by matching the experimentally measured parameters $\als(M_{\ssst{Z}})$, $G_F$, $M_W$, $M_Z$, $M_{\ssst{t}}$ and $M_{\ssst{H}}$ to them \cite{Buttazzo:2013uya,Kniehl:2015nwa},
where the uncertainties of $G_F$, $M_W$ and $M_Z$ have a negligible influence on the $\overline{\text{MS}}$-couplings as compared to the other three. The Higgs mass is taken to be
\mbox{$M_{\ssst{H}}= 125.09\pm 0.24$ GeV} \cite{Aad:2015zhl} and the strong coupling is extracted from \mbox{$\als(M_{\ssst{Z}})=0.1181\pm 0.0013$} \cite{pdg2015}.
 
The top mass is a more difficult subject. At the moment a theoretically well-defined mass is not available to high prescision. The extraction of an $\overline{\text{MS}}$-mass from cross section measurements
leads to an uncertainty of $4-5$ GeV \cite{pdg2015}, the extraction of the pole mass from cross section measurements to \cite{pdg2015} 
\be M_t=174.6^{\ssst{pole}} \pm 1.9 \text{ GeV.} \label{pdgtoppolemass} \ee
The most precise top mass measurements from LHC and TEVATRON give \mbox{$M_{\ssst{t}}^{\ssst{MC}} \approx 173.34\pm 0.76$ GeV} \cite{ATLAS:2014wva} where
the Monte Carlo mass parameter $M_{\ssst{t}}^{\ssst{MC}}$ would correspond to the pole mass in a purely perturbative setup. Since $M_{\ssst{t}}^{\ssst{MC}}$ is affected by real emission and the implementation
of the parton shower it eludes a theoretical definition based on the Lagrangian but has to be calibrated to fit the experimental data. The pole mass of a heavy quark also suffers from a conceptual problem,
namely the renormalon ambiguity \cite{Bigi:1994em,Beneke:1998ui}. 
This is due to the fact that a quark is not observed as a free particle and non-perturbative effects spoil the convergence of the relations between
the pole mass and e.~g.~an $\overline{\text{MS}}$-mass. One way to connect the Monte Carlo and the pole mass is to identify the first with a theoretically well-defined short-distance mass at a scale 
$R\sim 1-3$ GeV, i.~e.~the parton shower cut-off of the Monte Carlo simulation \cite{Hoang:2008yj}. This so-called MSR mass does not suffer from non-perturbative effects such as renormalons 
due to the IR cutoff. The MSR mass can in turn be connected to the pole mass (by adding the contributions from the IR region) \cite{Moch:2014tta} leading to 
\mbox{$M_{\ssst{t}}^{\ssst{pole}} = 173.39\pm 0.76 \text{(exp)} {}^{+0.82}_{-0.62}\text{(th)}$ GeV} for 
\mbox{$M_{\ssst{t}}^{\ssst{MC}} = M_{\ssst{t}}^{\ssst{MSR}}(3^{+6}_{-2}) \text{GeV}$} \cite{Moch:2014lka} or combining these errors in quadrature
\be M_{\ssst{t}}^{\ssst{pole}} = 173.39 {}^{+1.12}_{-0.98} \text{ GeV.} \label{mochtoppolemass}\ee 
For the following analysis we take this value which is very close to the Monte Carlo mass,
noting however that the discrepancy with the current PDG top pole mass value \eqref{pdgtoppolemass} indicates that the uncertainties on this parameter might be even larger and a direct and precise extraction of
theoretically well-defined quantities like the MSR mass or the $\overline{\text{MS}}$ top-Yukawa coupling itself from the experimental data of a linear collider 
will be necessary in order to
match the size of the other uncertainties entering the vacuum stability analysis.

For \mbox{$M_{\ssst{t}} = 173.39 {}^{+1.12}_{-0.98}$ GeV} \cite{ATLAS:2014wva,Moch:2014lka}
we get the couplings in the $\overline{\text{MS}}$-scheme at the scale of the top mass using two-loop matching relations \cite{Buttazzo:2013uya}
\bea 
\gs(M_{\ssst{t}})&=&1.1652 \pm 0.0035 \text{(exp)}, \nonumber \\
\yt(M_{\ssst{t}})&=&0.9374  {}^{+0.0063}_{-0.0062}\text{(exp)} \pm 0.0005 \text{ (2 loop matching)}, \nonumber \\
\lambda(M_{\ssst{t}})&=&0.1259 \pm 0.0005 \text{(exp)} \pm 0.0003 \text{ (2 loop matching)}, \label{initialcond}\\
\gw(M_{\ssst{t}})&=& 0.6483,\nonumber \\
\gb(M_{\ssst{t}})&=& 0.3587,\nonumber
\eea
where the experimental uncertainty (exp) stems from $M_t, M_H$ and $\als(M_Z)$ and the theoretical one (theo) from the matching of 
on-shell to \msbar{} parameters \cite{Buttazzo:2013uya}. Evaluating $\beta_{\sss{\yt}}$ at the scale $M_{\ssst{t}}$ we find
\bea
\f{\beta_{\sss{\yt}}^{(1)}(\mu=M_{\ssst{t}})}{(16\pi^2)} &=& -2.4 \times 10^{-2}, \nonumber \\
\f{\beta_{\sss{\yt}}^{(2)}(\mu=M_{\ssst{t}})}{(16\pi^2)^2} &=& -2.9 \times 10^{-3}, \nonumber \\
\f{\beta_{\sss{\yt}}^{(3)}(\mu=M_{\ssst{t}})}{(16\pi^2)^3} &=& -1.2 \times 10^{-4}, \\
\f{\beta_{\sss{\yt}}^{(4)}(\mu=M_{\ssst{t}})}{(16\pi^2)^4} &=& +5.9 \times 10^{-6}, \nonumber 
\eea
which shows the expected suppression of higher order contributions. The four-loop terms are significantly smaller than the lower loop terms.
For $\beta_{\sss{\lambda}}$ however the picture is different. Already in previous works \cite{Chetyrkin:2012rz,Chetyrkin:2013wya,Chetyrkin:2013wyaERR} we found a slow
convergence of the perturbation series up to three-loop order at the electroweak scale and this is also found true at four-loop level:
\bea
\f{\beta_{\sss{\lambda}}^{(1)}(\mu=M_{\ssst{t}})}{(16\pi^2)} &=& -1.0 \times 10^{-2}, \nonumber \\
\f{\beta_{\sss{\lambda}}^{(2)}(\mu=M_{\ssst{t}})}{(16\pi^2)^2} &=& -2.3 \times 10^{-5}, \nonumber \\
\f{\beta_{\sss{\lambda}}^{(3)}(\mu=M_{\ssst{t}})}{(16\pi^2)^3} &=& +1.1 \times 10^{-5}, \\
\f{\beta_{\sss{\lambda}}^{(4)}(\mu=M_){\ssst{t}}}{(16\pi^2)^4} &=& +1.3 \times 10^{-5}. \nonumber 
\eea
As discussed in \cite{Chetyrkin:2013wya,Chetyrkin:2013wyaERR}
there is a remarkable cancellation between the terms containing only $\gs$ and $\yt$ at two-loop\footnote{Here for example the numerically largest terms 
$\propto \yt^4 \gs^2$, $\propto \yt^6$ and $\propto \lambda \yt^2 \gs^2$ cancel so well that the sum of these terms is only about $2-3\%$ of the size of the largest
term $\propto \yt^4 \gs^2$ at $\mu=M_{\ssst{t}}$.} and three-loop level at the scale of the top mass making the overall contribution much smaller than the 
size of the individual terms. These cancellations seem to accidental, so we cannot predict there existence at four-loop level.
However, if the cancellation also exists at this order the terms $\propto \yt^6\gs^4,\,\propto \yt^8\gs^2,\,\propto \yt^{10},$ etc could make the result significantly smaller and hence 
increase the convergence of the perturbative series.
At higher scales, where $\gs$ and $\yt$ become smaller the convergence is of course better also if we only include the term $\propto \yt^4\gs^6$. 
At $\mu=10^{9}\text{GeV}$ we find
\bea
\f{\beta_{\sss{\lambda}}^{(1)}(\mu=10^{9}\text{GeV})}{(16\pi^2)} &=& -1.4 \times 10^{-3}, \nonumber \\
\f{\beta_{\sss{\lambda}}^{(2)}(\mu=10^{9}\text{GeV})}{(16\pi^2)^2} &=& +8.0 \times 10^{-8}, \nonumber \\
\f{\beta_{\sss{\lambda}}^{(3)}(\mu=10^{9}\text{GeV})}{(16\pi^2)^3} &=& +3.1 \times 10^{-7}, \\
\f{\beta_{\sss{\lambda}}^{(4)}(\mu=10^{9}\text{GeV})}{(16\pi^2)^4} &=& +6.9 \times 10^{-8}, \nonumber 
\eea
where the leading four-loop contribution is a factor $4-5$ smaller than the three-loop result but still not completely negligible.
We will now check what this means for the evolution of the Higgs self-coupling $\lambda$ up to the Planck scale. 

\section{Evolution of $\lambda$ and vacuum stability}

\begin{figure}[!ht] \begin{center}
\includegraphics[width=0.99\linewidth]{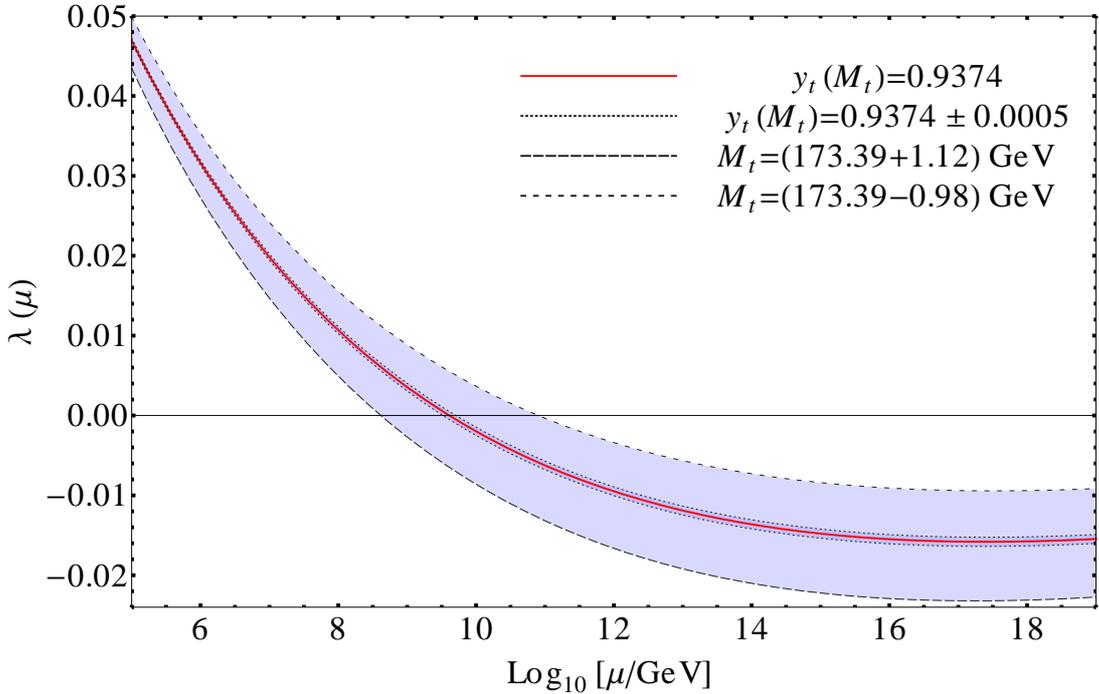} \end{center}
\caption{Evolution of $\lambda$: top matching and top measurement uncertainties for $M_t=173.39$ GeV.} \label{lambda_uncert} 
\end{figure}

We evolve $\lambda$ from the scale $M_{\ssst{t}}$ using the initial conditions \eqref{initialcond} and the full SM $\beta$-functions (including $\gs,\yt,\gw,\gb,\lambda$) up
to three-loop order and at four-loop level $\beta^{(4)}_{\gs}(\gs,\yt,\lambda)$ \cite{Bednyakov:2015ooa,Zoller:2015tha} and the leading contributions to 
$\beta^{(4)}_{\yt}(\gs,\yt)$ and $\beta^{(4)}_{\lambda}(\gs,\yt)$, as given in \eqref{4lbetayt} and \eqref{4lbetalambda}.
Fig.~\ref{lambda_uncert} shows the result compared to the largest remaining uncertainties on the theory and on the experimental side.
The smaller error band marks the $1\sigma$ uncertainty stemming from the top matching, i.e. we vary $\yt$ by $\pm 0.0005$ (see \eqref{initialcond}).
The larger error band marks the $1\sigma$ uncertainty stemming from the top pole mass.

\begin{figure}[!ht] \begin{center}
\includegraphics[width=0.99\linewidth]{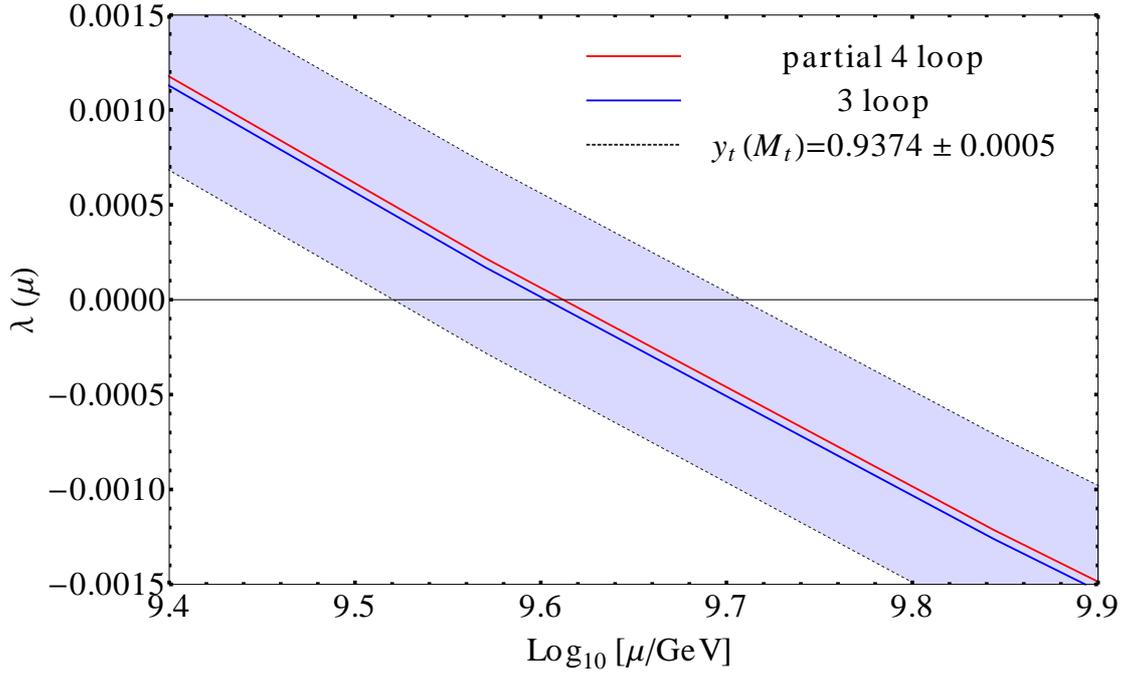} \end{center}
\caption{Evolution of $\lambda$: 3 loop and partial 4 loop $\beta$-functions, top matching uncertainty.} \label{lambda_34} 
\end{figure}

The difference between the evolution of $\lambda$ with three-loop $\beta$-functions (blue curve) and including the leading
four-loop terms (red curve) should give some indication on the uncertainty stemming from the truncation of the perturbative series for the $\beta$-functions.
In order to see this difference we have to zoom in. We choose to do this at the scale where $\lambda$ becomes negative, which is shown in Fig.~\ref{lambda_34}.
The error bands are calculated using the partial four-loop results and hence centered around the red curve.

\begin{figure}[!ht] \begin{center}
\includegraphics[width=0.99\linewidth]{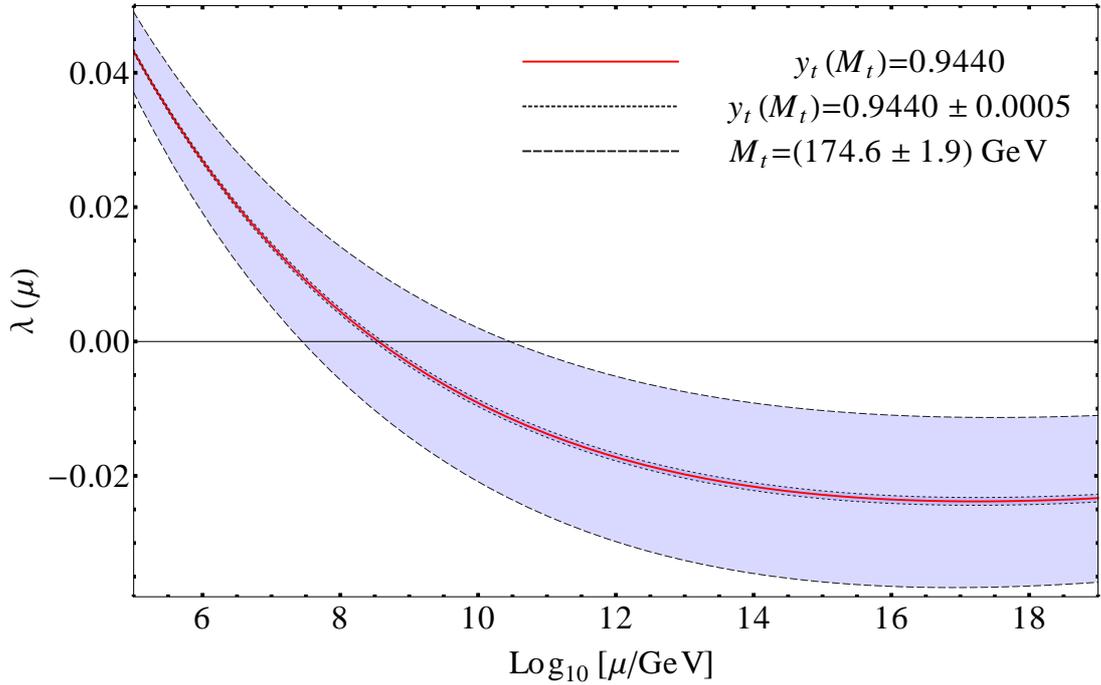} \end{center}
\caption{Evolution of $\lambda$: top matching and top measurement uncertainties for $M_t=174.6$ GeV.} \label{lambda_uncert_pdg} 
\end{figure}

In order to further illustrate the dependence of the vacuum stability problem on the top mass and the importance of the issues related to the top pole mass as an input parameter
we show the evolution of $\lambda$ also for the PDG value extracted from cross section measurements of the top pole mass \eqref{pdgtoppolemass} and the corresponding uncertainties in Fig.~\ref{lambda_uncert_pdg}.

The conclusion for vacuum stability remains the same as in our previous works \cite{Chetyrkin:2012rz,Zoller:2012cv,
Zoller:2014cka,Zoller:2014xoa,Zoller:2013mra}. It looks as if $\lambda$ becomes negative at around $\log_{10}\lb\f{\mu}{\text{GeV}}\rb\sim 9.6$ (or $\log_{10}\lb\f{\mu}{\text{GeV}}\rb\sim 8.7$ 
in Fig.~\ref{lambda_uncert_pdg}) rendering the SM not stable if extended up to scales above, but a definitive answer is pending on a more precise extraction of $\yt(\mu_0)$ from experimental data.
It is worth noting, however, that due to the reduction in the top mass uncertainty since the combined LHC and TEVATRON analysis \cite{ATLAS:2014wva} a stable
SM up to the Planck scale is strongly disfavoured. 

\section{Conclusions \label{last}}

In this work we have presented analytical results for the leading four-loop contributions to
the $\beta$-function for the Higgs self-coupling $\lambda$ and the top-Yukawa coupling as well as to the anomalous dimension of the Higgs field. These
results have been connected to pure QCD RG functions and a relation between the anmalous dimension $\gamma_0$ of the vacuum energy and $\gamma^{SS}_m$ was found.
We have performed an analysis of the evolution of the Higgs self-coupling updating the analyses presented in previous works \cite{Chetyrkin:2012rz,Zoller:2012cv,
Zoller:2014cka,Zoller:2014xoa,Zoller:2013mra} and establishing a nice hirarchy between the different sources of uncertainty. 

With the computation of the leading four-loop terms to
$\beta_{\lambda}, \beta_{\yt}$ and $\beta_{\gs}$ the $\beta$-function uncertainty to the question of vacuum stability becomes significantly smaller
than the matching uncertainty (before the two were comparable) which is in turn significantly smaller than the experimental top mass uncertainty.
We expect that a full calculation of four-loop $\beta$-functions in the SM will confirm this conclusion 
rendering the remaining $\beta$-function uncertainty almost negligible in comparison to the other sources
of uncertainty for a vacuum stability analysis.

\section*{Acknowledgements}

One of us (K.~Ch.) thanks  Mikhail Shaposhnikov for a hint about a possible
connection between the vacuum energy and $\beta_{\lambda}$.

This research was supported in part by the Swiss National Science Foundation (SNF) under contract BSCGI0\_157722.
The work by K.~Ch.~was supported by the Deutsche Forschungsgemeinschaft through CH1479/1-1.

\bibliographystyle{JHEP}

\bibliography{LiteraturSM}

\end{document}